\documentstyle[12pt,epsfig]{article}
\def\lapproxeq{\lower .7ex\hbox{$\;\stackrel{\textstyle
<}{\sim}\;$}}
\def\gapproxeq{\lower .7ex\hbox{$\;\stackrel{\textstyle
>}{\sim}\;$}}
\begin{document}
%
\pagestyle{empty}
\begin{center}
{\large\bf  Spin dependent structure function $g_1$ at
small $x$  and  small $Q^2$}
\footnote{Contribution to the 3{rd} UK Phenomenology Workshop on
HERA Physics, St. John's College, Durham, UK, September 1998}
\vspace{1.1cm}\\
         {\sc B.~Bade\l{}ek $^a$} and {\sc J.~Kwieci\'nski $^b$} \\
\vspace{0.3cm}
$^a$ {\it Department of Physics, Uppsala University, P.O.Box 530,
751 21 Uppsala, Sweden} \\
{\it and Institute of Experimental Physics, Warsaw University, Ho\.za 69,
00-681 Warsaw, Poland (e-mail: badelek@tsl.uu.se)}\\

$^b$ {\it Department of Theoretical Physics, 
H.~Niewodnicza\'nski Institute of Nuclear Physics, \\
Radzikowskiego 152, 31-342 Cracow, Poland (e-mail:
jkwiecin@solaris.ifj.edu.pl)} \\
\end{center}
\vskip5mm
The SMC has recently published the optimal set of results
for the virtual photon-proton
and virtual photon-deuteron cross section asymmetries
$A_1^{\rm p}(x,Q^2)$ and $A_1^{\rm d}(x,Q^2)$, \cite{optimal},
obtained from spin dependent inclusive muon--proton and
muon--deuteron scattering at 100 and 190 GeV.
The data covered the kinematic interval of 0.0008$<$$x$$<$0.7
and 0.2$<$Q$^2$$<$100 GeV$^2$. Events with lower values of $x$
were not measured to avoid a contamination with muon scattering off
atomic electrons at $x=$0.000545. Recently the SMC has obtained results
for the $A_1^{\rm p}$ and $A_1^{\rm d}$ at very small $x$ and $Q^2$,
acquired from the 190 GeV $\mu$-$p$ and $\mu$-$d$ scattering, 
\cite{t15}, Fig.1.
The data were collected with
a de\-di\-ca\-ted "low $x$ trigger" in which both a minimal energy deposit
in the hadronic part of the calorimeter and a
scattered muon were demanded. This together with off-line selections
removed all but (5$\pm$1)$\% ~\mu e$ events and considerably reduced the
radiative background. Measured asymmetries are very small, cf.Fig.1, and
special care has to be taken to well control the systematic errors, in
particular those due to uncertainties of the spin averaged structure
functions $R$ and $F_2$ at low $x$ and low $Q^2$. 
The new data complement the optimal set
of the SMC results in the region of 0.01 $< Q^2 <$ 0.2 GeV$^2$ 
and 0.00006 $< x < $0.0008 and include the lowest values of $x$ ever
measured for the spin dependent inelastic scattering. 


Below we shall review the predictions concerning the $g_1(x,Q^2)$ valid
in the region of the new measurements. 
To obtain predictions for the asymmetry $A_1^p$ rather
than for $g_1^p$, $A_1\approx 2x(1+R)g_1/F_2$,
models for $F_2^p$ and $R$,
valid at low $x$ and $Q^2$, like e.g. these of ref. \cite{BBJK2}
may be used.

The $g_1$ function is expected to be a finite 
function of $W^2$ in the limit  $Q^2 \rightarrow 0$ for 
fixed $W^2$,  free
from any kinematical singularities or zeros at $Q^2=0$,
similarily to the structure function $F_1$. 

The new SMC data include the kinematic region where
the four momentum transfer
is smaller than any other energy scale, $W^2$ or $2M\nu$,
$Q^2\ll W^2$ and $W^2$ is high, $W^2\gapproxeq$ 100 GeV$^2$.
Thus one should expect that the Regge model should be applicable there.  
This model gives the $x$ (or $W^2$) dependence of $g_1$ at fixed $Q^2$.
However $W^2$ changes very little in the kinematic range of the data:
from about 100 GeV$^2$ at $x=$0.1 to about 220 GeV$^2$ at $x=$0.0001,
contrary to a quite strong change of $Q^2$ (from about 20 GeV$^2$ 
to about 0.01 GeV$^2$ respectively).  
This means that the new SMC measurements cannot test the Regge 
behaviour of $g_1$ through the $x$ dependence of the latter. 
For completeness we shall only list the Regge predictions. 
According to the Regge model $g_1(x,Q^2) \sim x^{-\alpha}$ for 
$x \rightarrow 0$ (for fixed $Q^2$),   where 
$\alpha$  denotes the intercept of the Regge pole trajectory corresponding 
to axial vector mesons.  It is expected that $\alpha \sim 0$ for both 
$I=0$ and $I=1$ trajectories, \cite{hei}. 
This behaviour of the $g_1$ should go smoothly to the $W^{2 \alpha}$  
dependence for $Q^2 \rightarrow 0$. 
Other considerations related to the Regge theory
predict  $g_1^p\sim${ln}$x$, \cite{clo_rob}, 
while the model based on exchange of two nonperturbative gluons gives 
$g_1^p\sim$ 2~{ln}(1/$x$)--1, \cite{bass_land}. A perverse behaviour,
$g_1\sim$1/$(x$ln$^2x)$, recalled in \cite{clo_rob}, is not valid  
for $g_1$, \cite{misha}.
%
%
%
%
\begin{figure}[ht]
\begin{center}
\vspace*{-0.5cm}
\hspace*{-0.5cm}
\epsfig{file=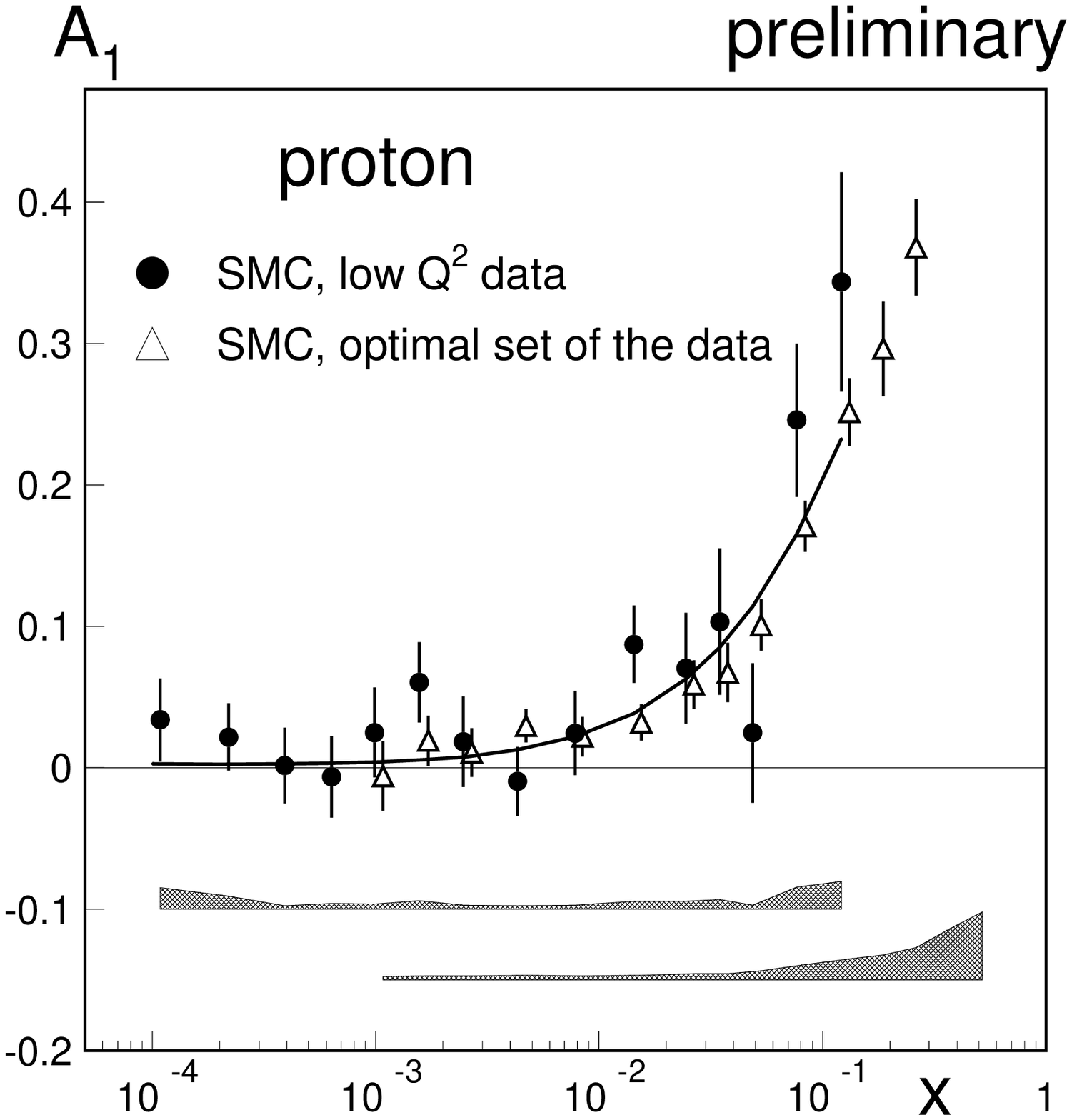,width=13cm}
\end{center}
{\footnotesize
Figure 1. $A_1^p$ as a function of $x$ measured by SMC.
Filled circles mark the preliminary new data extending to very low $x$, 
\protect\cite{t15}, triangles 
-- that of ref. \protect\cite{optimal}. Errors are statistical.
Systematic ones are marked as shaded bands.
The curve corresponds to the calculations based on eqs. (\ref{dpi},\ref{gp1}). 
}
\end{figure}
%
%
%
%
\\

 In perturbative QCD the structure function $g_1$
is controlled at low $x$ by the {\it double} logarithmic
ln$^2(1/x)$ contributions  
\cite{BARTSNS}.  
It is convenient to discuss the ln$^2(1/x)$ resummation
using the formalism of the   unintegrated (spin dependent) parton
distributions
$f_i(x^{\prime},k^2)$ ($i=u_{v},d_{v},\bar u,\bar d,\bar s,g$) where
$k^2$
is the transverse momentum squared of the parton $i$ and $x^{\prime}$
the
longitudinal momentum fraction of the parent nucleon carried by a parton
\cite{BBJK,BZIAJA}.
The conventional (integrated) distributions $\Delta p_i(x,Q^2)$ are
related in the
following way  to the unintegrated distributions $f_i(x^{\prime},k^2)$:
\begin{equation}
 \Delta p_i(x,Q^2)=\Delta p_i^{(0)}(x)+
 \int_{k_0^2}^{W^2}{dk^2\over k^2}f_i(x^{\prime}=
x(1+{k^2\over Q^2}),k^2)
\label{dpi}
\end{equation}
where  $W^2=Q^2(1/x-1)$ and $\Delta p_i^{(0)}(x)$ denote the
nonperturbative
part of the
of the distributions.  The $g_1^p$ is related in a standard
way to the (integrated) quark and antiquark distributions, i.e.  \\
$$
g_1^p(x,Q^2)=
$$
\begin{equation}
{1\over 2}\left[{4\over 9}(\Delta u_v(x,Q^2) + 2\Delta \bar u (x,Q^2))+
{1\over 9}(\Delta d_v(x,Q^2) + 2\Delta \bar u(x,Q^2) +  2\Delta \bar
s(x,Q^2))
\right]
\label{gp1}
\end{equation}
where $\Delta u_v(x,Q^2) = \Delta p_{u_v}(x,Q^2)$ etc.
We assume $\Delta \bar u =\Delta \bar d$ and 
number of flavours $N_F$=3.
The parameter $k_0^2$ is the infrared cut-off
($k_0^2 \sim $1 GeV$^2$).

The sum of ln$^2(1/x)$
terms is  generated by the corresponding integral equations
for
the functions  
$f_i((x^{\prime},k^2)$ \cite{BBJK,BZIAJA,MANAR}. 
%
These equations lead to 
approximate $x^{-\lambda}$ behaviour of the $g_1$
with $\lambda \sim 0.3$ and 
$\lambda \sim 1$ for the non-singlet and singlet parts 
respectively  
which is more singular at low $x$ than 
that   
generated by the (non-perturbative) 
Regge pole exchanges.  
The ln$^2(1/x)$ effects are presumably not important
in the $W^2$ range of the fixed target experiments, cf.Fig.2 in
\cite{BBJK}, but
they significantly affect $g_1$ in the low $x$ region which may 
be probed at the polarized HERA, \cite{BBJK,BZIAJA}.

The formalism based on the 
unintegrated distributions is very suitable for extrapolating 
$g_1$ to the 
region of low $Q^2$ at fixed $W^2$ \cite{BBJK}. 
Since $x(1+k^2/Q^2) \rightarrow k^2/W^2$ for $Q^2 \rightarrow 0$  in the integrand 
in eq. (\ref{dpi}) and since $k^2 >k_0^2$,  
the $g_1(x,Q^2)$ defined by eqs. (\ref{dpi},\ref{gp1}) 
can be smoothly extrapolated to 
$Q^2=0$    provided that
$\Delta p_i^{(0)}(x)$ are free from
kinematical singularities at $x=0$, as in parametrisations used  in refs.
\cite{BBJK,BZIAJA} where $\Delta p_i^{(0)}(x) = C_i
(1-x)^{n_i}$. 
If $\Delta p_i^{(0)}(x)$ contain kinematical singularities
at $x=0$ then
one may replace  $\Delta p_i^{(0)}(x)$
with $\Delta p_i^{(0)}(\bar x)$ where
$\bar x = x\left(1+{k_0^2/ Q^2}\right)$
and leave remaining parts of the calculation unchanged.
However
the (extrapolated)  partonic contribution
to the low $Q^2$ region may not be the only one there; 
the VMD part may play non-neglible role as well. 
Predictions based on equations (\ref{dpi},\ref{gp1}) 
shown as a curve in Fig.1 reproduce a general trend in the data.
They are systematically smaller than the measurements.
There is thus a room for non-partonic contributions to $g_1^p$.

\end{document}